\documentclass{article}

\usepackage[numbers]{natbib}

     \usepackage[preprint]{tackling_climate_workshop_style}

\usepackage[utf8]{inputenc} % allow utf-8 input
\usepackage[T1]{fontenc}    % use 8-bit T1 fonts
\usepackage{hyperref}       % hyperlinks
\usepackage{url}            % simple URL typesetting
\usepackage{booktabs}       % professional-quality tables
\usepackage{amsfonts}       % blackboard math symbols
\usepackage{nicefrac}       % compact symbols for 1/2, etc.
\usepackage{microtype}      % microtypography
\usepackage{amsmath}
\usepackage{graphicx}

\title{Data Assimilation using ERA5, ASOS, and the
U-STN model for Weather Forecasting over the UK}

\author{Wenqi Wang\\
	Department of Earth Science and Engineering\\
	Imperial College London\\
	\texttt{wenqi.wang21@imperial.ac.uk} \\
 	\And
	Jacob Bieker\\
	  Open Climate Fix\\
	\texttt{jacob@openclimatefix.org} \\
	\And
	Rossella Arcucci\\
	Department of Earth Science and Engineering\\
	Imperial College London\\
	\texttt{r.arcucci@imperial.ac.uk} \\
        \And	
        César Quilodrán-Casas\\
	Department of Earth Science and Engineering\\
	Imperial College London\\
	\texttt{c.quilodran@imperial.ac.uk} \\
 }
\hypersetup{
pdftitle={A template for the arxiv style},
pdfsubject={q-bio.NC, q-bio.QM},
pdfauthor={David S.~Hippocampus, Elias D.~Striatum},
pdfkeywords={First keyword, Second keyword, More},
}

\begin{document}
\maketitle

\begin{abstract}
In recent years, the convergence of data-driven machine learning models with Data Assimilation (DA) offers a promising avenue for enhancing weather forecasting. This study delves into this emerging trend, presenting our methodologies and outcomes. We harnessed the UK's local \verb+ERA5+ 850 hPa temperature data and refined the U-STN12 global weather forecasting model, tailoring its predictions to the UK's climate nuances. From the \verb+ASOS+ network, we sourced \verb+T2m+ data, representing ground observations across the UK. We employed the advanced kriging method with a polynomial drift term for consistent spatial resolution. Furthermore, Gaussian noise was superimposed on the \verb+ERA5+ \verb+T850+ data, setting the stage for ensuing multi-time step synthetic observations. Probing into the assimilation impacts, the \verb+ASOS+ \verb+T2m+ data was integrated with the \verb+ERA5+ \verb+T850+ dataset. Our insights reveal that while global forecast models can adapt to specific regions, incorporating atmospheric data in DA significantly bolsters model accuracy. Conversely, the direct assimilation of surface temperature data tends to mitigate this enhancement, tempering the model's predictive prowess.
\end{abstract}

\section{Introduction}
Numerical Weather Prediction (NWP) has progressed over the decades, establishing itself as a cornerstone in weather forecasting~\cite{weyn_can_2019}. Its efficacy, demonstrated across diverse environmental scenarios, provides probabilistic forecasts~\cite{palmer_ecmwf_2019}. Among various NWP models, the European Centre for Medium-Range Weather Forecasting (ECMWF) IFS (Integrated Forecasting System) is consistently recognised for its accurate representation of atmospheric conditions in designated geographical areas~\cite{rasp_data-driven_2021}. While it maintains a lead over machine learning (ML) models, the rise of ML-driven models — with their refined designs and increasing resolutions — cannot be overlooked~\cite{rasp2023weatherbench}. These models leverage historical atmospheric data, aspiring to match or surpass IFS predictions while conserving computational resources and boosting processing speeds. This includes resolutions from 5.625 degrees in the WeatherBench Convolution Neural Network (CNN) series~\cite{rasp_weatherbench_2020} to 1.0 degree in Keisler's Graph Neural Network (GNN) model~\cite{keisler_forecasting_2022} and 0.25 degree in GraphCast~\cite{lam_graphcast_2022}. Even though ML model developers do not anticipate replacing NWP entirely, the strides made in ML suggest its capacity to refine existing forecasting approaches~\cite{lam_graphcast_2022}. In recent years, the integration of data-driven ML models and Data Assimilation (DA) techniques has attracted substantial attention due to its potential to enhance model performance~\cite{buehner_ensemble_2017}. DA serves as an intermediary, merging observational data with model outputs to hone initial backgrounds, thereby fostering improved forecast results. Integrating obervational data in a NWP is not trivial as the model needs to be restarted from new initial conditions~\cite{casas2020reduced}. Integrating new observational data in an ML model through fine-tuning via DA has been shown to be very effective~\cite{cheng2023machine}. In the domain of atmospheric research, 850 hPa temperature (\verb+T850+) and 500 hPa geopotential height (\verb+Z500+) are fundamental datasets. Their significance stems from their representation of large-scale circulation in the troposphere, which inherently affects near-surface weather conditions and extremities~\cite{chattopadhyay_towards_2022}. The objective of weather forecasting is to predict the future state of the atmosphere for a designated time~\cite{rasp_weatherbench_2020}. However, this forecasting is confined to the atmosphere's prediction horizon, estimated to be approximately two weeks~\cite{scher2019generalization}.

In this study, we not only used data from ERA5 and the ASOS network but also incorporated global meteorological data from the U-STN model. ML models excel at handling heterogeneous data from multiple sources. They can automatically identify correlations between different datasets and effectively integrate this information, which is crucial for improving DA outcomes.
Weather systems are highly complex and nonlinear. ML models, especially deep learning models, are adept at capturing this complexity, providing a finer understanding compared to traditional numerical weather prediction models. One key advantage of ML models is their ability to adapt and respond to changes in new data. This means that as we collect more meteorological data and real-time observations, the models can continuously self-optimize and adjust, thereby enhancing long-term prediction accuracy.

\section{Methods}
\label{sec:headings}

In this study, we adapted a global weather forecasting model~\cite{chattopadhyay_towards_2022} employing a U-Net~\cite{ronneberger2015u} enhanced with a deep spatial transformer (U-STN) and focused on a single data variable. This model was specifically retrained for the UK region. To explore the optimisation effects of DA  on predictive outcomes, we incorporated the sigma-point Ensemble Kalman Filter (SPEnKF) algorithm~\cite{noauthor_sigma-point_nodate}, particularly leveraging surface data from ground observation stations. A primary model was trained using a distinct 12-hour time interval on \verb+ERA5+ dataset: U-STN12. In our study, four different data typologies were utilized for DA:
(1) Employing $\sigma_{\text{obs}} = \sigma_z$ and $\sigma_{\text{obs}} = 0.5\sigma_z$, the \verb+T850+ data was augmented with Gaussian noise at two levels, simulating observational data with noise coefficients of 1 and 0.5, respectively. This methodology is aligned with standard practices in DA and facilitates the examination of DA's impact~\cite{chattopadhyay_towards_2022}; (2) Utilizing $\sigma_{\text{obs}} = \sigma_z$ and $\sigma_{\text{obs}} = 0.5\sigma_z$, synthetic observations were generated on \verb+T850+ via U-STN12 model, offering a closer approximation to actual observations than Gaussian noise; (3) Observational data were directly simulated using data from ASOS observation stations; (4)The \verb+ERA5+ \verb+T2m+ data served as a surrogate for simulated observational data. Theoretically, its data distribution mirrors that of the ASOS data, thereby mitigating potential alterations in data distribution characteristics during the processing of ASOS data. Our primary objective is to find the influence of incorporating ground observational data in DA processes for atmospheric datasets.

\subsection{Datasets}
\subsubsection{ERA5}
The ECMWF's \verb+ERA5+ reanalysis dataset integrates NWP forecast models with contemporaneous observations using 4D-Var DA~\cite{h_era5_nodate}. We collected \verb+T850+ hourly data from 1940 to present~\cite{bell_era5_2021} and \verb+T2m+ hourly data from 1940 to present, encompassing the timeframe from 1979 to 2022. The 2022 annual \verb+T850+ temperature pattern is shown in \ref{fig: map} (b). For more details of the data and the processing procedures, see Appendix \ref{sec:appendix}
\subsubsection{ASOS Dataset}
\begin{figure}[h]
    \begin{center}
        \includegraphics[width=1\textwidth]{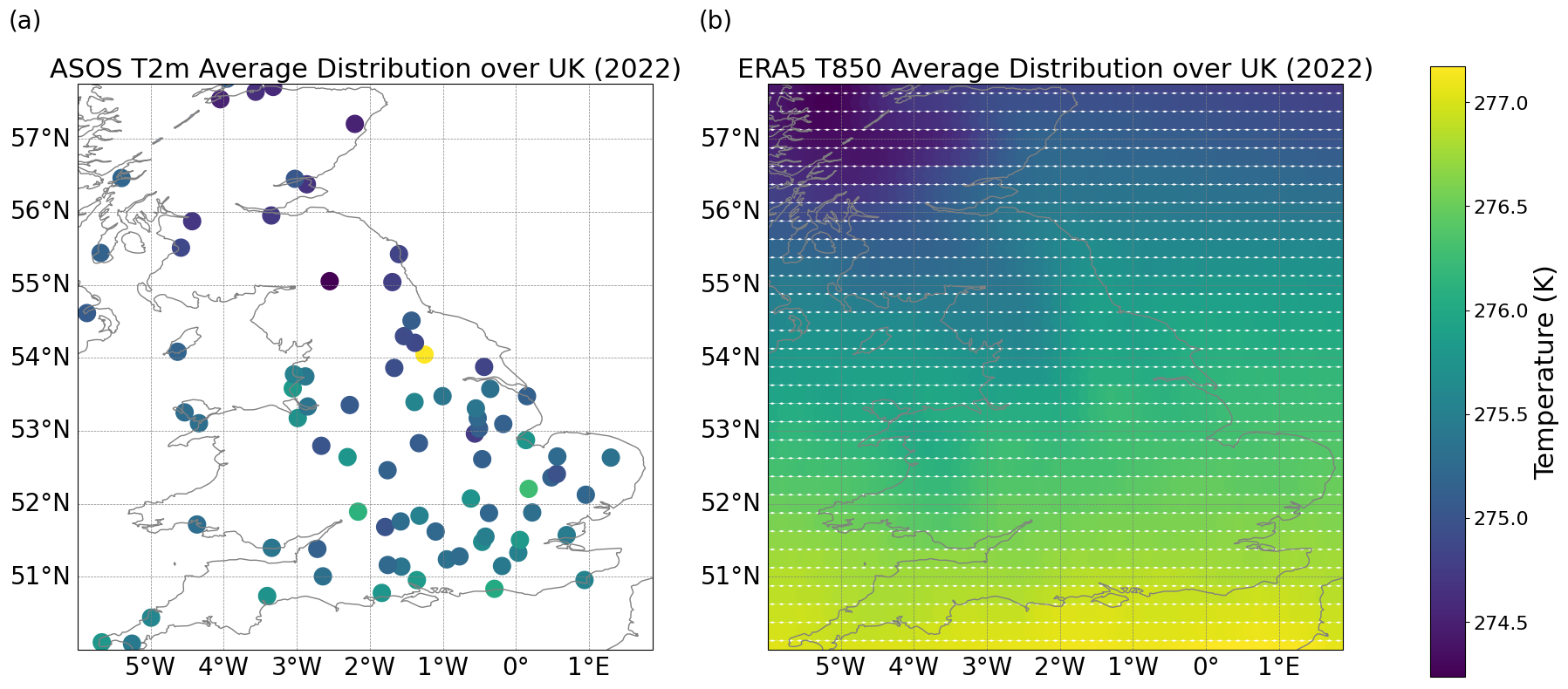}
    \end{center}
    \caption{(a) Geographical distribution of ASOS surface observation stations and annual average T2m values between latitudes 50° to 57.75° and longitudes -6° to 1.875°(2022). (b) Annual average T850 temperature pattern from ERA5 within the same latitudinal and longitudinal bounds (2022).}
    \label{fig: map} 
    
\end{figure}

Recognising the potential correlations with surface observational data, we incorporated 2-meter temperature (\verb+T2m+) data from \verb+ASOS+ via the Iowa Environmental Mesonet ~\cite{herzmann_iem_download_2023}, which delivers hourly surface observations spanning the period from 1979 to 2022 and encapsulating readings from 112 observation stations within the UK. This system predominantly supports aviation weather forecasting and further weather forecasting research ~\cite{lengel_everything_2018}. The 2022 annual \verb+T2m+ temperature and distribution of observation stations are depicted in \ref{fig: map} (a). Our aim with the \verb+ASOS+ data was to align its structure with the \verb+ERA5+ data, as visualised in Figure \ref{fig: map}. 

 % Figure

\subsection{Model}

We employed a U-Net-based ML model enhanced with a deep spatial transformer~\cite{chattopadhyay_towards_2022}. Using the Adam optimiser~\cite{kingma_adam_2017} and a mean squared error (MSE) loss function, the model's parameters were aligned with those specified in ~\cite{chattopadhyay_towards_2022}. Although our primary goal is the exploration of DA from ground observation stations, promising results may prompt further hyperparameter optimisation specific to the UK or other regions. Our prediction approach is autoregressive: a prediction at time $T(t+ \Delta t)$ becomes the input for $T(t+ 2\Delta t)$. Depending on the forecast's intended duration, $\Delta t$ might vary. In this work, we set $\Delta t$ at 12h for a model named U-STN12. The specific model code and implementation are available on GitHub: \url{https://anonymous.4open.science/r/TacklingClimateChangeAIforNeurIPS2023_2_02DF/src/models}.

\subsection{Data assimilation}

We adopted the SPEnKF approach~\cite{merwe_sigma-point_2004}. The SPEnKF, unlike the EnKF, employs an unscented transformation that utilizes sigma points, which are a predefined set of points used to estimate the mean and variance of a nonlinear function.  By processing these sigma points through the nonlinear function, it becomes feasible to gauge the output's mean and covariance without relying on perturbation-based linearisation~\cite{noauthor_sigma-point_nodate}. The specific assimilation code and implementation are available on GitHub: \url{https://github.com/acse-ww721/DA_ML_ERA5_ASOS_Weather_Forecasting_UK}.

\section{Results}
\label{sec:results}
\begin{figure*}[h]
    \begin{center}
        \includegraphics[width=1\textwidth]{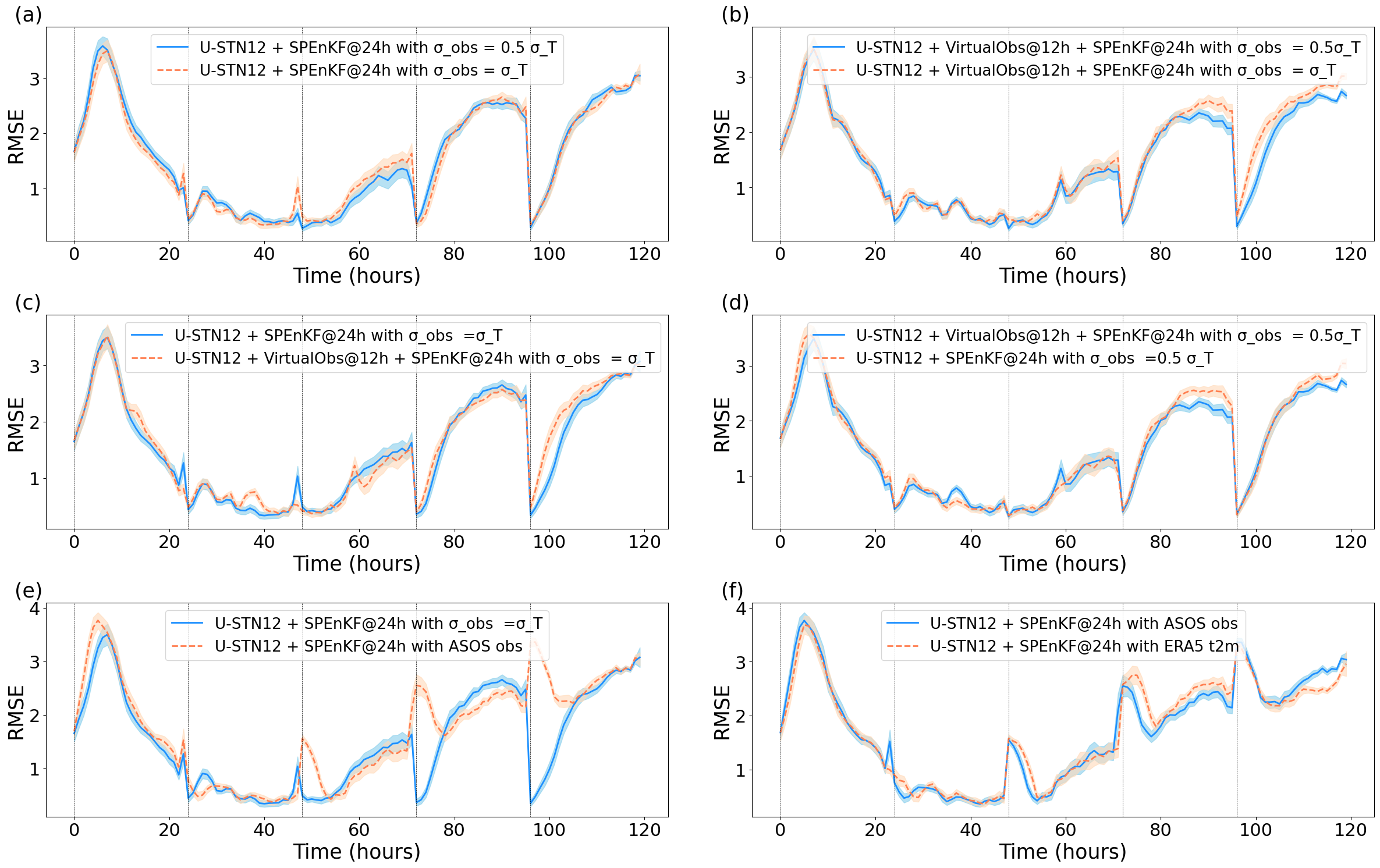}
    \end{center}
    \caption{\label{fig: experiment} Root Mean Square Error (RMSE) Comparison for the U-STN12-SPEnKF Model Across Different DA Sources.}
\end{figure*}

 % Figure
We employed U-STN12 as our ML model and SPEnKF as the DA algorithm. The prediction's initial value is derived from a noisy observation. Every 24 hours, we assimilate this noisy observation to refine the prediction. By introducing various noise levels, Gaussian noise $N(0,\sigma_{\text{obs}})$ is superimposed onto the \verb+ERA5+ \verb+T850+ data. We then analysed the root mean square error (RMSE) between the predicted mean and the noisy data in the \verb+T850+ domain over the initial 120 hours across 50 random conditions. The result is shown in Figure \ref{fig: experiment}. Our attention is primarily drawn to the similarities and disparities in trends before and after the DA juncture. Given that our model commences training from the 12 lead time, the RMSE for the initial 12 hours exhibits an upward trend due to lack of training. Between the 12 and 48 hours lead time, there is a consistent reduction in RMSE, underscoring the model's effective predictive performance. Notably, a marked decrease in RMSE is observed at the 24-hour lead time point, corresponding with the introduction of DA. This decrease is followed by a gradual increase in RMSE over time, likely a consequence of accumulating losses. A scenario where $\sigma_{obs} = \sigma_{T}$ results in a larger RMSE, suggesting that increased noise levels detrimentally impact prediction accuracy. Nonetheless, the RMSE trends remain consistent between $\sigma_{obs} = \sigma_{T}$ and $\sigma_{obs} = 0.5\sigma_{T}$, attesting to the model's robustness.\\
\\
As illustrated in Figure \ref{fig: experiment} (b), We experimented with using the model's training data at a time step $t=12$ hours as simulated observations, as an alternative to Gaussian noise simulated observations over a 24-hour DA span. Both $\sigma_{obs} = 0.5\sigma_{T}$ and  $\sigma_{obs} = \sigma_{T}$ exhibit consistent performance trends. Notably, with every 12-hour increment when synthetic observations are introduced, there is a discernible reduction in RMSE. \\
\\
In Figure \ref{fig: experiment} (c) and (d), when contrasting this with the RMSE trend without the addition of synthetic observations, it is evident that introducing these synthetic observations results in a more gradual ascent in the model's RMSE. When juxtaposed with the RMSE trajectory without synthetic observations, the early stages, characterised by high model prediction accuracy, did not manifest any pronounced benefits from the synthetic measurements. However, past the 48-hour mark, as the model's predictive efficacy weakens, the RMSE resulting from the inclusion of synthetic observations was notably lower than that from Gaussian noise simulation.\\
\\
As illustrated in Figure \ref{fig: experiment} (e), the integration of \verb+T2m+ data from the \verb+ASOS+ ground observation station into the DA process impairs its efficacy. A significant RMSE spike at the assimilation juncture manifestly evidences this degradation. As Figure \ref{fig: experiment} reveals, the original \verb+ASOS+ ground observation stations exhibit a sparse and non-uniform distribution. When interpolated to match the \verb+ERA5+ dataset resolution, inevitable errors emerge. To validate that discrepancies between ground and atmospheric data diminish DA performance, we similarly integrated \verb+ERA5+ \verb+T2m+ data into the assimilation process. As shown in Figure\ref{fig: experiment} (f), this integration mirrors the previous trend, with RMSE elevating at assimilation points, confirming the detrimental impact on DA. However, it is salient that the RMSE when utilising \verb+ASOS+ ground data is less than that from the \verb+ERA5+ \verb+T2m+ data. This observation paves the way for further optimisation of the interpolation technique and exploration of ground-atmospheric observation correlations.
% \subsection{Data assimilation}

\section{Conclusion}
\label{sec:conclusion}
Adapting a global ML model to regional predictions is viable.
However, regional-specific re-tuning of hyperparameters or model architecture is essential. The SPEnKF assimilation method, when applied with model-based noisy or multi-time step synthetic observations, augments prediction accuracy. Given the inherent discrepancies and spatial irregularities of \verb+ASOS+ observational data, especially within the UK context, the direct assimilation of surface observation data with atmospheric \verb+T850+ appears inadvisable. Based on our findings, we propose three refinements: (1) Data Sources and Interpolation: Acquire denser datasets and implement advanced interpolation and preprocessing techniques to better align \verb+T2m+ values with \verb+T850+; (2) Hyperparameter Tuning: Optimise hyperparameters in our models, including evaluating the number of autoregressive time steps, to enhance performance; and (3) Incorporation of Multi-Layer Pressure Levels: Integrate ground observation data with models employing multi-layer pressure levels, such as Graph Neural Networks (GNNs), to reduce interpolation errors and improve prediction accuracy. Through these enhancements, we aim to uncover patterns in using ground observation data for DA. Beyond the structural and data-based refinements previously mentioned, incorporating regional domain knowledge can dramatically improve performance and adaptability. Establishing a feedback system where the predictions of the model are continuously compared against actual observations. Any discrepancies can be used as learning points for the model, enabling it to self-correct and adapt to the specificities of the region over time.

\clearpage
% References
\bibliographystyle{plain}
\bibliography{references}  % BibTeX references are saved in references.bib

\clearpage
\section*{Appendix}
\label{sec:appendix}

\subsection*{Datasets}
\subsubsection*{ERA5}
We utilised the \verb+ERA5+ dataset spanning the years 1979 to 2020 for training, which comprises approximately 367,920 samples. The 2021 dataset, containing around 8,760 samples, was designated for validation, while the 2022 dataset, with a similar count of roughly 8,760 samples, was allocated for testing. Both \verb+ERA5+ datasets encompass a geographical expanse with latitude ranging from 49\textdegree to 58\textdegree and longitude from -8\textdegree to +2\textdegree.

The inception year of 1979 was strategically chosen, reflecting a pronounced enhancement in data availability and quality post this year. This period correlates with significant advancements in ancillary research domains, such as climate change studies~\cite{h_era5_nodate}.

To ensure integrity in the preprocessing of the downloaded raw data, efforts were directed to retain its intrinsic properties. This strategy aids subsequent comparisons with prevailing global weather forecast models~\cite{chattopadhyay_towards_2022} and the WeatherBench~\cite{rasp_weatherbench_2020}and WeatherBench2~\cite{rasp2023weatherbench}. The preprocessing steps involved:

(1) Cropping: The focus area was delimited to latitudes between 57.75\textdegree and 50.0\textdegree, and longitudes between -6° and 1.875\textdegree.
(2) Remeshing: The tool xsemf~\cite{zhuang_pangeo-data_xESMF_2023}was employed to realize a grid configuration of 32x64.
(3) Handling Null Values: In instances of null data points, a 3-hour sliding window was applied, imputing the average value.
Consequently, the annual datasets were re-gridded, conforming to the dimensions (time, latitude, longitude) of 8760×32×64, with a slight adjustment to 8784×32×64 for leap years.

\subsubsection*{ASOS Dataset}
We adopted the \verb+ASOS+ dataset by the following preprocessing steps:
(1) Hourly Normalisation: Recognizing the varied reporting schedules of \verb+ASOS+ stations, we standardized the hourly reporting times. Specifically: (a)For timestamps within the first 30 minutes of an hour (e.g., 01:20:00), if a full-hour data point precedes it, we omit the data. If not, it's shifted to the preceding full hour. (b)For timestamps beyond the 30-minute mark (e.g., 01:50:00), if a full-hour data point follows, the data is omitted; otherwise, it's shifted to the subsequent full hour.
(2)Region Filtering: We retained data within the bounds of latitude [50, 58] and longitude [-6, +2].
(3) Kriging Interpolation: Given the uneven station density, we used kriging interpolation via gstools~\cite{muller_geostat-framework/GSTools_2023} to ensure spatial consistency. Polynomial drift terms were incorporated to capture trends along latitude and longitude, and their interplay. The formula for drift is:

\begin{equation}
\text{drift} = F(1, \text{lat}, \text{lon}, \text{lat}^2, \text{lon}^2, \text{lat} \times \text{lon})
\end{equation}
where \(1\) is the baseline constant term, the \(\text{lat}\) term represents the linear variation with latitude, the \(\text{lon}\) term indicates the linear variation with longitude, \(\text{lat}^2\) denotes the quadratic trend with latitude, \(\text{lon}^2\) signifies the quadratic trend with longitude, and \(\text{lat} \times \text{lon}\) term captures the interaction of latitude and longitude.

In meteorology, spatial data often exhibit non-stationarity, meaning that the statistical characteristics of the data (such as mean and variance) vary across space. For example, the climate in the UK is influenced by various factors, including topography and proximity to the ocean, which can lead to different trends in variables like temperature and rainfall across space. The Kriging method with polynomial drift effectively handles this non-stationarity by introducing a polynomial trend term to capture these spatial variations.

 The polynomial drift term allows the model to consider trends in variables as they vary geographically. This is particularly important when dealing with variables like temperature, which may change with latitude, altitude, and other geographic factors. By modelling these trends, we can more accurately estimate meteorological conditions at unobserved points.





% \end{thebibliography}

\end{document}